\documentclass[preprint,12pt]{elsarticle}

\usepackage{graphicx} 	  
\usepackage{amssymb} 	

\usepackage{color}
\usepackage{amsmath}
\usepackage{mathtools}
\usepackage{fullpage}
\usepackage{algorithmic}

\algsetup{linenosize=\small}
\usepackage[table,xcdraw]{xcolor}
\usepackage{multirow}
\usepackage[super]{nth}
\usepackage{graphicx,adjustbox}
\usepackage{caption}
\usepackage[labelformat=simple]{subcaption}
\usepackage{comment}
\usepackage{setspace}
\usepackage{textcomp}
\usepackage{xspace}
\usepackage{siunitx}
\usepackage{epsfig}
\usepackage{epstopdf}
\usepackage{soul}
\usepackage{url}
\usepackage{tablefootnote}
\usepackage[linesnumbered,ruled]{algorithm2e}
\usepackage{booktabs}
\usepackage{hyperref}
\usepackage[normalem]{ulem}
\usepackage{footnote}
\usepackage[misc,geometry]{ifsym} 
\makesavenoteenv{tabular}
\usepackage[utf8]{inputenc}

\let\OldTexttrademark\texttrademark
\renewcommand{\texttrademark}{\OldTexttrademark\xspace}%

\usepackage{epsfig}
\usepackage{amssymb}

\journal{Ad-hoc Networks}

\begin{document}
	
	\begin{frontmatter}
		
		
		
		\title{Collaborative Spatial Reuse in Wireless Networks via Selfish Multi-Armed Bandits}
		
		\author[label1, label4]{Francesc Wilhelmi}\ead{francisco.wilhelmi@upf.edu}
		\author[label2]{Cristina~Cano}
		\author[label3]{Gergely~Neu}
		\author[label1]{Boris~Bellalta} 
		\author[label3]{Anders~Jonsson}
		\author[label1]{Sergio~Barrachina-Mu\~noz}
		\address[label1]{Wireless Networking Research Group (WN-UPF), 08018 Barcelona, Spain}
		\address[label2]{Wireless Networks Research Group (WINE-UOC), 08860 Castelldefels (Barelona), Spain}
		\address[label3]{Artificial Intelligence and Machine Learning Research Group (AIML-UPF), 08018 Barcelona, Spain}
		
		\begin{abstract}
			Next-generation wireless deployments are characterized by being dense and uncoordinated, which often leads to inefficient use of resources and poor performance. To solve this, we envision the utilization of completely decentralized mechanisms to enable Spatial Reuse (SR). \textcolor{black}{In particular, we focus on dynamic channel selection and Transmission Power Control (TPC). We rely on Reinforcement Learning (RL), and more specifically on Multi-Armed Bandits (MABs), to allow networks to learn their best configuration}. In this work, we study the exploration-exploitation trade-off by means of the $\varepsilon$-greedy, EXP3, UCB and Thompson sampling action-selection, and compare their performance. \textcolor{black}{In addition, we study the implications of selecting actions simultaneously in an adversarial setting \textcolor{black}{(i.e., concurrently)}, \textcolor{black}{and compare it with a sequential approach}.} Our results show that optimal proportional fairness can be achieved, even when no information about neighboring networks is available to the learners and Wireless Networks (WNs) operate selfishly. However, there is high temporal variability in the throughput experienced by the individual networks, especially for $\varepsilon$-greedy and EXP3. \textcolor{black}{These strategies, contrary to UCB and Thompson sampling, base their operation on the absolute experienced reward, rather than on its distribution.} We identify the cause of this variability to be the adversarial setting of our setup in which the set of most played actions provide intermittent good/poor performance depending on the neighboring decisions. \textcolor{black}{We also show that \textcolor{black}{learning sequentially, even if using a selfish strategy,} contributes to minimize this variability. \textcolor{black}{The sequential approach is therefore shown to effectively deal with the challenges posed by the adversarial settings that are typically found in decentralized WNs.}}
		\end{abstract}
		
		\begin{keyword}
			High-Density Wireless Networks; Spatial Reuse; Resource Allocation; Decentralized learning; Multi-Armed Bandits 
		\end{keyword}
		
	\end{frontmatter}
	
	
	\newpage
	
	\section{Introduction}
	\label{section:introduction}
	Due to the growing popularity of wireless deployments, especially the ones based on the IEEE 802.11 standard (i.e., Wi-Fi), it is very common to find independent overlapping Wireless Networks (WNs) sharing the same channel resources. \textcolor{black}{The decentralized nature of such kind of deployments leads to a significant lack of organization and/or agreement on sharing policies. As a result, resources are typically used inefficiently. An illustrative example of this can be found in \cite{akella2007self}, where the authors show that the power level used by wireless devices is typically set, by default, to the maximum, regardless of the distance between communicating nodes, and the channel occupancy. Consequently, increasing the capacity of such networks has become very challenging.}
	
	\textcolor{black}{Wireless networks operate in three main domains: time, frequency and space. While the first two have been largely exploited, the spatial domain still shows plenty of room for improvement. According to \cite{alawieh2009improving}, Spatial Reuse (SR) can be addressed by means of Transmission Power Control (TPC), Carrier Sense Threshold (CST) adjustment, rate adaptation (related to power control), and directional transmissions. In addition, interference cancellation can play a key role on spectral efficiency optimization \cite{miridakis2013survey}. On one side, TPC and CST adjustment aim at increasing spectral efficiency omnidirectionally. On the other hand, beamforming is meant for directional transmissions. Both beamforming and interference cancellation can be categorized as multiple antenna strategies. While the former allows to reduce the interference levels, the second one is useful to perform multiple simultaneous transmissions.}
	
	\textcolor{black}{In this work, we focus on Dynamic Channel Allocation (DCA) and TPC to address the decentralized SR problem. A proper frequency planning allows to reduce the interference between wireless devices, and tuning the transmit power adds an extra level of SR that can result in improved throughput and fairness. The application of TPC and DCA is particularly challenging by itself. The interactions among devices depend on many features (such as position, environment or transmit power) and are hard to derive. 
		Including beamforming and/or interference cancellation techniques \cite{dovelos2018breaking}, on the other hand, requires first a clear understanding of TCP and DCA performance alone, and is therefore left as future work.}
	
	\textcolor{black}{Motivated by these challenges,} we focus attention on Reinforcement Learning (RL), which has recently emerged as a very popular method to solve many well-known problems in wireless communications. \textcolor{black}{RL allows to reduce the complexity generated in wireless environments by finding practical solutions. By applying RL, optimal (or near-to-optimal) solutions can be obtained without having a full understanding on the problem in advance. So, one of the main goals of this paper is to show its feasibility for the decentralized SR problem.} Some RL-based applications can be found for packet routing \cite{littman1993distributed}, Access Point (AP) selection \cite{bojovic2011supervised, bojovic2012neural}, optimal rate sampling \cite{combes2014optimal}, or energy harvesting in heterogeneous networks \cite{miozzo2015distributed}. All these applications make use of online learning, where a learner (or agent) obtains data periodically and uses it to predict future good-performing actions. Online learning is particularly useful to cope with complex and dynamic environments. This background encourages us to approach a solution for the decentralized SR problem in WNs through online learning techniques.
	
	From the family of online algorithms, we \textcolor{black}{are interested on analyzing the performance of} Multi-Armed Bandits (MABs) \cite{BCB12} \textcolor{black}{when applied to WNs. The MAB model is well-known} in the online learning literature for solving resource allocation problems. In MABs, a given agent seeks to learn a hidden reward distribution while maximizing the gains. This is known as the exploration-exploitation trade-off. Exploitation is meant to maximize the long-term reward given the current estimate, and exploration aims to improve the estimate. Unlike classical RL, MABs do not consider states\footnote{A state refers to a particular situation experienced by a given agent, which is defined by a set of conditions. By having an accurate knowledge of its current situation, an agent can define state-specific strategies that maximize its profits.} in general, which can be hard to define for the decentralized SR problem presented in this work. On the one hand, spatial interference cannot be binary treated, thus leading to complex interactions among nodes. On the other hand, the adversarial setting unleashed by decentralized deployments increases the system complexity. Therefore, the obtained reward does not only depends on the actions taken by a given node, but also on the adversaries behavior.
	
	\textcolor{black}{This article extends our previous results presented in \cite{wilhelmi2017implications}. Here we generalize the contributions done by implementing several action-selection strategies to find the best combination of frequency channel and transmit power in WNs. These strategies are applied to the decentralized SR problem, where independent WNs learn selfishly, based on their own experienced performance.} On the one hand, we evaluate the impact of varying parameters intrinsic to the proposed algorithms on the resulting throughput and fairness. In addition, we analyze the effects of learning selfishly, and shed light on the future of decentralized approaches. Notably, we observe that even though players act selfishly, some of the algorithms learn to play actions that enhance the overall performance, some times at the cost of high temporal variability. Considering selfish WNs and still obtaining collaborative behaviors is appealing to typical chaotic and dynamic deployments. \textcolor{black}{Finally, the adversarial setting in WNs is studied under two learning implementations: namely \textit{concurrent} and \textit{sequential}. Both procedures rule the operation followed by learners (based on the proposed action-selection strategies). In particular, WNs select an action at the same time for the concurrent approach. In contrast, an ordered action-selection procedure is followed for the sequential case. We study the performance of the aforementioned techniques in terms of convergence speed, average throughput and variability.} The main contributions of this work are summarized below:
	\begin{itemize}		
		\item We devise the feasibility of applying MAB algorithms as defined in the online learning literature to solve the resource allocation problem in WNs.
		\item We study the impact of different parameters intrinsic to the action-selection strategies considered (e.g., exploration coefficients, learning rates) on network performance. \textcolor{black}{In addition, we analyze the implications derived from the application of different learning procedures, referred to as \textit{concurrent} and \textit{sequential}, which rule the moment at which WNs act.}
		\item \textcolor{black}{We show the impact of learning concurrently and sequentially. In particular, the former leads to a high throughput variability experienced by WNs, which is significantly reduced by the sequential approach.} \textcolor{black}{Accordingly, we envision the utilization of sequential approaches to achieve decentralized learning in adversarial wireless networks.}
		\item Finally, we show that there are algorithms that learn to play collaborative actions even though the WNs act selfishly, which is appealing to practical application in chaotic and dynamic environments. In addition, we shed light on the root causes of this phenomena.  
	\end{itemize}
	
	The remaining of this document is structured as follows: Section \ref{section:related_work} outlines relevant related work. Section \ref{section:mabs} introduces the proposed learning algorithms and their practical implementation for the resource allocation problem in WNs. Then, Section \ref{section:system_model} presents the simulation scenarios and the considerations taken into account. The simulation results are later presented in Section \ref{section:performance_evaluation}. Finally, Section \ref{section:conclusions} provides the final remarks.
	
	\section{Related Work}
	\label{section:related_work} 
	
	\textcolor{black}{Decentralized SR has been considerably studied by the wireless research community. The authors in \cite{argyriou2010collision} propose using relay nodes to re-transmit packets lost as a result of a collision. The relay node is able to decode different signals from the environment and to detect if a collision took place. Then, with the aim of improving the re-transmissions operation, it benefits from the current transmission to forward the decoded packets to their original destinations. Although this method shows performance improvements in dense scenarios where collisions are very likely to occur, its effectiveness is subject to the network topology. Regarding directional transmissions, the authors in \cite{babich2015design} propose two novel access schemes to allow multiple simultaneous transmissions. In particular, nodes' activity information is sensed, which, together with antenna's directionality information, allows to build a new set of channel access rules.}
	
	\textcolor{black}{Despite approaches based on directional transmissions and interference cancellation are very powerful and allow to significantly increase SR, they strongly rely on having multiple antennas. Such a requirement is not mandatory for the SR operation based on TPC and CST adjustment. In this work, we focus on the former because tuning the transmit power has a direct impact on the generated interference. This allows to purely study the interactions that occur among nodes implementing decentralized SR. Moreover, we consider DCA to be combined with TPC, so that further potential gains can be achieved.}
	
	\textcolor{black}{DCA has been extensively studied from the centralized perspective, especially through techniques based on graph coloring \cite{riihijarvi2005frequency, mishra2005weighted}. Despite these kind of approaches allow to effectively reduce the interference between WNs, a certain degree of communication is required. Regarding decentralized methods, the authors in \cite{akl2007dynamic} propose a very simple approach in which each AP maintains an interference map of their neighbors, so that channel assignment is done through interference minimization. Unfortunately, the interactions among APs in the decentralized setting are not studied. Separately, \cite{chen2007improved} proposes two decentralized approaches that rely on the interference measured at both APs and stations (STAs) to calculate the best frequency channels for dynamic channel allocation. To do so, a WN, in addition to the interference sensed by its associated devices, considers other metrics such as the amount of traffic, so that some coordination is required at the neighbor level (e.g., periodic reporting). The authors in \cite{yue2011cacao} show that the decentralized DCA problem is NP-hard. In addition, they propose a distributed algorithm whereby APs select the best channel according to the observed traffic information (i.e., channel sensing is considered).}
	
	\textcolor{black}{In this work we aim to extend the approach in \cite{yue2011cacao} in two ways. \textcolor{black}{First, we aim to provide a flexible solution based on the performance achieved by a given WN.} Second, we aim to tackle the spatial domain through TPC\textcolor{black}{, which has been shown to provide large improvements in wireless networks \cite{elbatt2000power}}. \textcolor{black}{However}, dealing with the spatial dimension leads to unpredictable interactions in terms of interference. \textcolor{black}{Such a complexity is illustrated} in \cite{tang2014joint}, which performs power control and rate adaptation in subgroups of Wireless Local Area Networks (WLANs). The creation of clusters allows defining independent power levels between devices in the same group, which are useful to avoid asymmetric links. However, to represent all the possible combinations, graphs can become very large, especially in high-density deployments. \textcolor{black}{When it comes to decentralized mechanisms, we find the work in \cite{gandarillas2014dynamic}, which applies TPC based on real-time channel measurements \cite{gandarillas2014dynamic}}. The proposed mechanism (so called Dynamic Transmission Power Control) is based on a set of triggered thresholds that increase/decrease the transmit power according to the state of the system. The main problem is that thresholds are set empirically (based on simulations), which limits the potential of the mechanisms in front of multiple scenarios.}
	
	\textcolor{black}{\textcolor{black}{As shown by previous research, the optimal decentralized SR in WNs through TPC and DCA is very hard to be derived analytically, mostly because of the adversarial setting and the lack of information at nodes. The existing decentralized solutions barely provide flexibility with respect to the scenario, so that potential use cases are disregarded.} For that, we focus on online learning, and more precisely Multi-Armed Bandits (MABs). The MAB framework allows to reduce the complexity of the SR problem, since detailed information about the scenario is not considered. In contrast, learners gain knowledge on all the adversaries as a whole, thus facing a single environment.} To the best of our knowledge, there is very little related work on applying MAB techniques to the problem of resource allocation in WNs. In \cite{coucheney2015multi}, the authors propose modeling a resource allocation problem in Long Term Evolution (LTE) networks through MABs. In particular, a set of Base Stations (BS) learn the best configuration of Resource Blocks (RBs) in a decentralized way. For that purpose, a variation of EXP3 (so-called Q-EXP3) is proposed, which is shown to reduce the strategy set. Despite a regret bound is provided, it is subject to the fact that an optimal resource allocation exists, i.e., every BS obtains the necessary resources. In addition, a large number of iterations is required to find the optimal solution in a relatively small scenario, thus revealing the difficulties shown by decentralized settings.
	
	More related to the problem proposed here, the authors in \cite{maghsudi2015joint} show a channel selection and power control approach in infrastructureless networks, which is modeled through bandits. In particular, two different strategies are provided to improve the performance of two Device to Device (D2D) users (each one composed by a transmitter and a receiver), which must learn the best channel and transmit power to be selected. Similarly to our problem, users do not have any knowledge on the channel or the other's configuration, so they rely on the experienced performance in order to find the best configuration. An extension of \cite{maghsudi2015joint} is provided by the same authors in \cite{maghsudi2015channel}, which includes a calibrated predictor (referred in the work as \textit{forecaster}) to infer the behavior of the other devices in order to counter act their actions. In each agent, the information of the forecaster is used to choose the highest-rewarding action with a certain probability, while the rest of actions are randomly selected. Henceforth, assuming that all the networks use a given strategy $\mathcal{X}$, fast convergence is ensured. Results show that channel resources are optimally distributed in a very short time frame through a fully decentralized algorithm that does not require any kind of coordination. Both aforementioned works rely on the existence of a unique Nash Equilibrium, which favors convergence. In contrast, in this article we aim to extend Bandits utilization to denser deployments, and, what is more important, to scenarios with limited available resources in which there is not a unique Nash Equilibrium (NE) that allows fast-convergence. Thus, we aim to capture the effects of applying selfish strategies in a decentralized way (i.e., agent $i$ follows a strategy $\mathcal{X}_i$ that does not consider the strategies of the others) and we also provide insight about the importance of past information for learning in dense WNs, which has not been studied before.
	
	\section{Multi-Armed Bandits for Improving Spatial Reuse in WNs}
	\label{section:mabs}
	
	In this work, we address the decentralized SR problem through online learning because of the uncertainty generated in an adversarial setting. The practical application of MABs in WNs is detailed next:
	
	\subsection{The Multi-Armed Bandits Framework}
	In the online learning literature, several MAB settings have been considered such as stochastic bandits \cite{thompson1933likelihood,lai1985asymptotically,auer2002finite}, adversarial bandits \cite{auer1995gambling,auer2002nonstochastic}, restless bandits \cite{whittle1988restless}, contextual bandits \cite{LCLS10} and linear bandits \cite{abe2003reinforcement,APS11}, and numerous exploration-exploitation strategies have been proposed such as \textit{$\varepsilon$-greedy} \cite{sutton1998reinforcement,auer2002finite}, \textit{upper confidence bound} (UCB) \cite{lai1985asymptotically,Agr95,BuKa96,auer2002finite}, \textit{exponential weight algorithm for exploration and exploitation} (EXP3) \cite{auer1995gambling,auer2002finite} and \textit{Thompson sampling} \cite{thompson1933likelihood}. The classical multi-armed bandit problem models a sequential interaction scheme between a learner and an environment. The learner sequentially selects one out of $K$ actions (often called \emph{arms} in this context) and earns some rewards determined by the chosen action and also influenced by the environment. Formally, the problem 
	is defined as a repeated game where the following steps are repeated in each round $t=1,2,\dots,T$:
	\begin{enumerate}
		\item The environment fixes an assignment of rewards $r_{a,t}$ for each action $a\in[K] \stackrel{\text{def}}{=} \left\{1,2,\dots,K\right\}$,
		\item the learner chooses action $a_t\in[K]$,
		\item the learner obtains and observes reward $r_{a_t,t}$.
	\end{enumerate}
	The bandit literature largely focuses on the perspective of the learner with the objective of coming up with learning algorithms that attempt to maximize the sum of the rewards gathered during the whole procedure (either with finite or infinite horizon). As noted above, this problem has been studied under various assumptions made on the environment and the structure of the arms. The most important basic cases are the \emph{stochastic} bandit problem where, for each particular arm $a$, the rewards are i.i.d.~realizations of random variables from a fixed (but unknown) distribution $\nu_a$, and the \emph{non-stochastic} (or \emph{adversarial}) bandit problem where the rewards are chosen arbitrarily by the environment. In both cases, the main challenge for the learner is the \emph{partial observability} of the rewards: the learner only gets to observe the reward associated with the chosen action $a_t$, but never observes the rewards realized for the other actions.
	
	Let ${\rm r}_{a^*,t}$ and ${\rm r}_{a,t}$ be the rewards obtained at time $t$ from choosing actions $a^*$ (optimal) and $a$, respectively. Then, the performance of learning algorithms is typically measured by the \emph{total expected regret} defined as  \[R_T = \sum_{t=0}^{T} \mathbb{E}\left[\left({\rm r}_{a^*,t} - {\rm r}_{a,t}\right)\right].\]
	
	An algorithm is said to \emph{learn} if it guarantees that the regret grows sublinearly in $T$, that is, if $R_T = o(T)$ is guaranteed as $T$ grows large, or, equivalently, that the average regret $R_T/T$ converges to zero. Intuitively, sublinear regret means that the learner eventually identifies the action with the highest long-term payoff. Note, as well, that the optimal action $a^*$ is the same across all the rounds. Most bandit algorithms come with some sort of a guaranteed upper bound on $R_T$ which allows for a principled comparison between various methods. 
	
	\subsection{Multi-Armed Bandits Formulation for Decentralized Spatial Reuse}
	
	\textcolor{black}{We model the decentralized SR problem through adversarial bandits. In such a model, the reward experienced by a given agent (WN) is influenced by the whole action profile, i.e., the configurations used by other competing WNs. From a decentralized perspective, the adversarial setting poses several challenges with respect to the existence of a NE. Ideally, the problem is solved if all the competitors implement a pure strategy\footnote{A pure strategy NE is conformed by a set of strategies and payoffs, so that no player can obtain further benefits by deviating from its strategy.} that allows maximizing a certain performance metric. However, finding such a strategy may not be possible in unplanned deployments, due to the competition among nodes and the scarcity of the available resources. Understanding the implications derived from such an adversarial setting in the absence of a NE is one the main goals of this paper, which, to the best of our knowledge, has been barely considered in the previous literature.}
	
	\textcolor{black}{In particular,} we model this adversarial problem as follows. Let arm $a \in \mathcal{A}$ (we denote the size of $\mathcal{A}$ with K) be a configuration \textcolor{black}{in terms of channel and transmit power (e.g., $a_1$ = \{Channel: 1, TPC: -15 dBm\})}. Let $\Gamma_{i,t}$ be the throughput experienced by $\text{WN}_i$ at time $t$, and $\Gamma_{i}^*$ \textcolor{black}{the optimal throughput.}\footnote{\textcolor{black}{The optimal throughput is achieved in case of isolation (i.e., when no interference is experienced in the selected channel).}} We then define the reward $r_{i,t}$ experienced by $\text{WN}_i$ at time $t$ as:
	
	\begin{equation}
	r_{i,t} = {\frac{\Gamma_{i,t}}{{\Gamma_{i}^*}}} \leq 1,
	\label{eq:reward_generation}
	\nonumber
	\end{equation}
	
	In order to attempt to maximize the reward, we have considered the \emph{$\varepsilon$-greedy}, \emph{EXP3}, \emph{UCB} and \emph{Thompson sampling} action-selection strategies, which are described next in this section. While $\varepsilon$-greedy and EXP3 explicitly include the concepts of \emph{exploration coefficient} and \emph{learning rate}, respectively, UCB and Thompson sampling are parameter-free policies that extend the concept of exploration (actions are explored according to their estimated value and not by commitment). \textcolor{black}{The aforementioned policies are widely spread and considered of remarkable importance in the MAB literature.} 
	
	\subsubsection{$\varepsilon$-greedy}
	\label{section:bandits_egreedy}	
	The \emph{$\varepsilon$-greedy} policy \cite{sutton1998reinforcement,auer2002finite} is arguably the simplest learning algorithm attempting to deal with exploration-exploitation trade-offs. In each round $t$, the $\varepsilon$-greedy algorithm explicitly decides whether to explore or exploit: with probability $\varepsilon$, the algorithm picks an arm uniformly at random (exploration), and otherwise it plays the arm with the highest empirical return $\hat{r}_{k,t}$ (exploitation). 
	
	In case $\varepsilon$ is fixed for the entire process, the expected regret is obviously going to grow linearly as $\Omega\left(\varepsilon T\right)$ in general. Therefore, in order to obtain a sublinear regret guarantee (and thus an asymptotically optimal growth rate for the total rewards), it is critical to properly adjust the exploration coefficient. Thus, in our \textcolor{black}{$\varepsilon$-greedy} implementation, we use a time-dependent exploration rate of $\varepsilon_t = \varepsilon_0 / \sqrt{t}$, as suggested in the literature \cite{auer2002finite}. The adaptation of this policy to our setting is shown as Algorithm \ref{alg:egreedy}.
	
	\begin{algorithm}[H]	
		\SetAlgoLined
		\KwIn{SNR: information about the Signal-to-Noise Ratio received at the STA, $\mathcal{A}$: set of possible actions in \{$a_1, ..., a_K$\}}
		\textbf{Initialize:} $t=0$, $\varepsilon_t = \varepsilon_0$, $r_{k} = 0, \forall a_k \in \mathcal{A}$\\
		\While{active}{
			Select $a_k$  $\begin{cases}
			\underset{k=1,...,K}{\text{argmax }} r_{k,t},& \text{with prob. } 1 - \varepsilon\\
			k \sim \mathcal{U}(1, K),              & \text{otherwise}
			\end{cases}$\\
			Observe the throughput experienced $\Gamma_t$\\
			Compute the reward $r_{k,t} = \frac{\Gamma_t}{\Gamma^*}$, where $\Gamma^* = B \log_{2}(1+\text{SNR})$ \\
			$\varepsilon_t \gets \varepsilon_0 / \sqrt{t}$ \\
			$ t \gets t + 1$
		}
		\caption{Implementation of Multi-Armed Bandits ($\varepsilon$-greedy) in a WN. $\mathcal{U}(1, K)$ is a uniform distribution that randomly chooses from 1 to $K$.}
		\label{alg:egreedy}			
	\end{algorithm}
	
	\subsubsection{EXP3}
	\label{section:bandits_exp3}	
	The EXP3 algorithm \cite{auer1995gambling,auer2002nonstochastic} is an adaptation of the weighted majority algorithm of \cite{LW94,FS97} to the non-stochastic bandit problem. EXP3 maintains a set of non-negative weights assigned to each arm and picks the actions randomly with a probability proportional to their respective weights (initialized to 1 for all arms). The aim of EXP3 is to provide higher weights to the best actions as the learning procedure proceeds. 
	
	More formally, letting $w_{k,t}$ be the weight of arm $k$ at time $t \in \{1,2 ...\}$, EXP3 computes the probability $p_{k,t}$ of choosing arm $k$ in round $t$ as
	\begin{equation}
	\label{eq:exp3_prob}
	p_{k,t} = (1-\gamma) \frac{w_{k,t}}{\sum_{i=1}^{\text{K}}w_{i,t}} + \frac{\gamma}{K},
	\nonumber
	\end{equation}
	where $\gamma\in[0,1]$ is a parameter controlling the rate of exploration.
	Having selected arm $a_t$, the learner observes the generated pay-off $r_{a_t,t}$ and computes the importance-weighted reward estimates for all $k\in[K]$
	\begin{equation}
	\label{eq:exp3_estimated_reward}
	\widehat{r}_{k,t} = \frac{\mathbb{I}_{\left\{I_t = k\right\}}r_{k,t}}{p_{k,t}},
	\nonumber
	\end{equation}
	where $\mathbb{I}_{\left\{A \right\}}$ denoting the indicator function of the event $A$ taking a value of $1$ if $A$ is true and $0$ otherwise.
	Finally, the weight of arm $k$ is updated as a function of the estimated reward:
	\begin{equation}
	\label{eq:exp3_weights}
	w_{k,t+1}=w_{k,t} e^{\frac{\eta \cdot \widehat{r}_{k,t}}{K}},
	\nonumber
	\end{equation}	
	where $\eta>0$ is a parameter of the algorithm often called the \emph{learning rate}. Intuitively, $\eta$ regulates the rate in which the algorithm incorporates new observations. \textcolor{black}{Large values of $\eta$ correspond to more confident updates and small values lead to more conservative behaviors}. As we did for the exploration coefficient in \emph{$\varepsilon$-greedy}, we use a time-dependent learning rate of $\eta_t = \eta_0 / \sqrt{t}$ \cite{auer2002finite}. Our implementation of EXP3 is detailed in Algorithm \ref{alg:exp3}.
	
	\begin{algorithm}[H]	
		\SetAlgoLined
		\KwIn{SNR: information about the Signal-to-Noise Ratio received at the STA, $\mathcal{A}$: set of possible actions in \{$a_1, ..., a_K$\}}
		\textbf{Initialize:} $t=0$, $\eta_t = \eta_0$, $w_{k,t} = 1, \forall a_k \in \mathcal{A}$\\
		\While{active}{
			$p_{k,t} \leftarrow (1-\gamma) \frac{w_{k,t}}{\sum^K_{i=1} w_{i,t}} + \frac{\gamma}{K}$ \\
			Draw $a_k \sim p_{k,t} = (p_{1,t}, p_{2,t}, ... , p_{K,t})$\\
			Observe the throughput experienced $\Gamma_t$\\
			Compute the reward $r_{k,t} = \frac{\Gamma_t}{\Gamma^*}$, where $\Gamma^* = B \log_{2}(1+\text{SNR})$ \\
			$\widehat{r}_{k,t} \leftarrow \frac{r_{k,t}}{p_{k,t}}$ \\
			$w_{k,t} \leftarrow w_{k,t-1}^{\frac{\eta_{t}}{\eta_{t-1}}} \cdot e^{\eta_{t} \cdot \widehat{r}_{k,t}}$\\
			$w_{k',t} \leftarrow w_{k',t-1}^{\eta_t / \eta_{t-1}}, \forall k' \neq k$\\
			$\eta_{t} \leftarrow \frac{\eta_0}{\sqrt{t}}$\\
			$t \leftarrow t + 1$\\
		}		
		\caption{Implementation of Multi-Armed Bandits (EXP3) in a WN}
		\label{alg:exp3}	
	\end{algorithm}
	
	\subsubsection{UCB}
	\label{section:bandits_ucb}		
	The \emph{upper confidence bound} (UCB) action-selection strategy \cite{Agr95,BuKa96,auer2002finite} is based on the principle of \emph{optimism in face of uncertainty}: in each round, UCB selects the arm with the highest statistically feasible mean reward given the past observations. Statistical feasibility here is represented by an upper confidence bound on the mean rewards which shrinks around the empirical rewards as the number of observations increases. Intuitively, UCB trades off exploration and exploitation very effectively, as upon every time a suboptimal arm is chosen, the corresponding confidence bound will shrink significantly, thus quickly decreasing the probability of drawing this arm in the future. The width of the confidence intervals is chosen carefully so that the true best arm never gets discarded accidentally by the algorithm, yet suboptimal arms are drawn as few times as possible. To obtain the first estimates, each arm is played once at the initialization. 
	
	Formally, let $n_k$ be the number of times that arm $k$ has been played, and $\Gamma_{k,t}$ the throughput obtained by playing arm $k$ at time $t$. The average reward $\overline{r}_{k,t}$ \textcolor{black}{of} arm $k$ at time $t$ is therefore given by:
	\begin{equation}
	\label{eq:ucb}
	\overline{r}_{k,t} = \frac{1}{n_k} \sum_{s=1}^{n_k} r_{k,s}
	\nonumber
	\end{equation}	
	Based on these average rewards, UCB selects the action that maximizes $\overline{r}_{k,t} + \sqrt{\frac{2 \ln(t)}{n_k}}$. By doing so, UCB implicitly balances exploration and exploitation, as it focuses efforts on the arms that are $i)$ the most promising (with large estimated rewards) or $ii)$ not explored enough (with small $n_k$). Our implementation of UCB is detailed in Algorithm \ref{alg:ucb}.	
	
	\begin{algorithm}[]
		\SetAlgoLined
		\KwIn{SNR: information about the Signal-to-Noise Ratio received at the STA, $\mathcal{A}$: set of possible actions in \{$a_1, ..., a_K$\}}
		\textbf{Initialize:} $t=0$, play each arm $a_k \in \mathcal{A}$ once\\
		\While{active}{
			Draw $a_k = \underset{k=1,...,K}{\text{argmax }} \overline{r}_{k} + \sqrt{\frac{2 ln(t)}{n_{k}}} $ \\
			Observe the throughput experienced $\Gamma_t$\\			
			Compute the reward $r_{k,t} = \frac{\Gamma_t}{\Gamma^*}$, where $\Gamma^* = B \log_{2}(1+\text{SNR})$ \\
			$n_k \leftarrow n_k + 1$\\
			$ \overline{r}_{k} \leftarrow \frac{1}{n_{k}} \sum_{s=1}^{n_{k}} r_{k,s}$\\
			$t \leftarrow t + 1$
		}		
		\caption{Implementation of Multi-Armed Bandits (UCB) in a WN}
		\label{alg:ucb}			
	\end{algorithm}	
	
	\subsubsection{Thompson sampling}
	\label{section:bandits_thompsons}	
	Thompson sampling \cite{thompson1933likelihood} is a well-studied action-selection technique that had been known for its excellent empirical performance \cite{CL11} and was recently proven to achieve strong performance guarantees, often better than those warranted by UCB \cite{AG12,KKM12,KKM13}. Thompson sampling is a Bayesian algorithm: it constructs a probabilistic model of the rewards and assumes a prior distribution of the parameters of said model. Given the data collected during the learning procedure, this policy keeps track of the posterior distribution of the rewards, and pulls arms randomly in a way that the drawing probability of each arm matches the probability of the particular arm being optimal. In practice, this is implemented by sampling the parameter corresponding to each arm from the posterior distribution, and pulling the arm yielding the maximal expected reward under the sampled parameter value.
	
	For the sake of practicality, we \textcolor{black}{assume that rewards follow a Gaussian distribution with a standard Gaussian prior,} as suggested in \cite{agrawal2013further}. By standard calculations, it can be verified that the posterior distribution of the rewards under this model is Gaussian with mean \textcolor{black}{and variance}
	\begin{equation}
	\textcolor{black}{\hat{r}_k(t) = \frac{\sum_{w=1:k}^{t-1} r_k(t) }{n_k(t) + 1} \Big/ \sigma_k^2(t) = \frac{1}{n_k + 1}},
	\nonumber
	\end{equation}
	where $n_k$ is the number of times that arm $k$ was drawn until the beginning of round $t$. Thus, implementing Thompson sampling in this model amounts to sampling a parameter $\theta_k$ from the Gaussian distribution $\mathcal{N}\left(\hat{r}_k(t),\sigma_k^2(t)\right)$ and choosing the action with the maximal parameter. Our implementation of Thompson sampling to the WN problem is detailed in Algorithm \ref{alg:thompsons}.
	
	\begin{algorithm}[]
		\SetAlgoLined
		\KwIn{SNR: information about the Signal-to-Noise Ratio received at the STA, $\mathcal{A}$: set of possible actions in \{$a_1, ..., a_K$\}}
		\textbf{Initialize:} $t=0$,  for each arm $a_k \in \mathcal{A}$, set $\hat{r}_{k} = 0$ and $n_k = 0$ \\
		\While{active}{
			For each arm $a_k \in \mathcal{A}$, sample $\theta_k(t)$ from normal distribution $\mathcal{N}(\hat{r}_{k}, \frac{1}{n_k + 1})$ \\
			Play arm $a_{k} = \underset{k=1,...,K}{\text{argmax }} \theta_k(t) $ \\
			Observe the throughput experienced $\Gamma_t$\\			
			Compute the reward $r_{k,t} = \frac{\Gamma_t}{\Gamma^*}$, where $\Gamma^* = B \log_{2}(1+\text{SNR})$ \\
			$ \hat{r}_{k,t} \leftarrow \frac{\hat{r}_{k,t}  n_{k,t} + r_{k,t}}{n_{k,t} + 2}$\\
			$n_{k,t} \leftarrow n_{k,t} + 1$\\
			$t \leftarrow t + 1$
		}		
		\caption{Implementation of Multi-Armed Bandits (Thompson \textcolor{black}{s.}) in a WN}
		\label{alg:thompsons}
	\end{algorithm}
	
	\section{System model}
	\label{section:system_model}	
	
	For the remainder of this work, we study the interactions among several WNs placed in a 3-D scenario that occur when applying MABs in a decentralized manner (with parameters described later in Section \ref{section:simulation_parameters}). For simplicity, we consider WNs to be composed by an AP transmitting to a single Station (STA) in a downlink manner. \textcolor{black}{Note that in typical uncoordinated wireless deployments (e.g., residential buildings), STAs are typically close to the AP to which they are associated. Thus, having several STAs associated to the same AP does not significantly impact the inter-WNs interference \textcolor{black}{studied} in this work.}
	
	\subsection{Channel modeling}
	\label{section:channel_modelling}		
	Path-loss and shadowing effects are modeled using the log-distance model for indoor communications. The path-loss between WN $i$ and WN $j$ is given by:	
	\begin{equation}
	\text{PL}_{i,j} = \text{P}_{{\rm tx},i} - \text{P}_{{\rm rx},j} = \text{PL}_0 + 10  \alpha  \log_{10}(d_{i,j}) + \text{G}_{{\rm s}} + \frac{d_{i,j}}{d_{{\rm obs}}} \text{G}_{{\rm o}}, \nonumber
	\nonumber
	\end{equation}
	where $\text{P}_{{\rm tx},i}$ is the transmitted power in dBm by the AP in $\text{WN}_i$, $\alpha$ is the path-loss exponent, $\text{P}_{{\rm rx},j}$ is the power in dBm received at the STA in $\text{WN}_j$, $\text{PL}_0$ is the path-loss at one meter in dB, $d_{i,j}$ is the distance between the transmitter and the receiver in meters, $\text{G}_{{\rm s}}$ is the log-normal shadowing loss in dB, and $\text{G}_{{\rm o}}$ is the obstacles loss in dB. Note that we include the factor $d_{{\rm obs}}$, which is the average distance between two obstacles in meters. 
	
	\subsection{Throughput calculation}
	\label{section:throughput_calculation}
	\textcolor{black}{The throughput experienced by WN $i$ at time $t$ is given by $\Gamma_{i,t} = B  \log_{2}(1 + \text{SINR}_{i, t})$,
		where $B$ is the channel width and SINR is the experienced Signal to Interference plus Noise Ratio. The latter is computed as $\text{SINR}_{i,t} = \frac{\text{P}_{i,t}}{\text{I}_{i,t}+\text{N}}$,
		where $\text{P}_{i,t}$ and $\text{I}_{i,t}$ are the received power and the sum of the interference at WN $i$ at time $t$, respectively, and N is the floor noise power.} Adjacent channel interference is also considered in $\text{I}_{i,t}$, so that the transmitted power leaked to adjacent channels is $20$ dBm lower for each extra channel separation. \textcolor{black}{Similarly, the optimal throughput is computed as $\Gamma_{i}^* = B  \log_{2}(1 + \text{SNR}_{i})$, which frames the operation of a given WN in isolation.}
	
	\subsection{Learning procedure}
	\label{section:learning_procedure}
	
\begin{figure}[h!]
		\centering				
		\epsfig{file=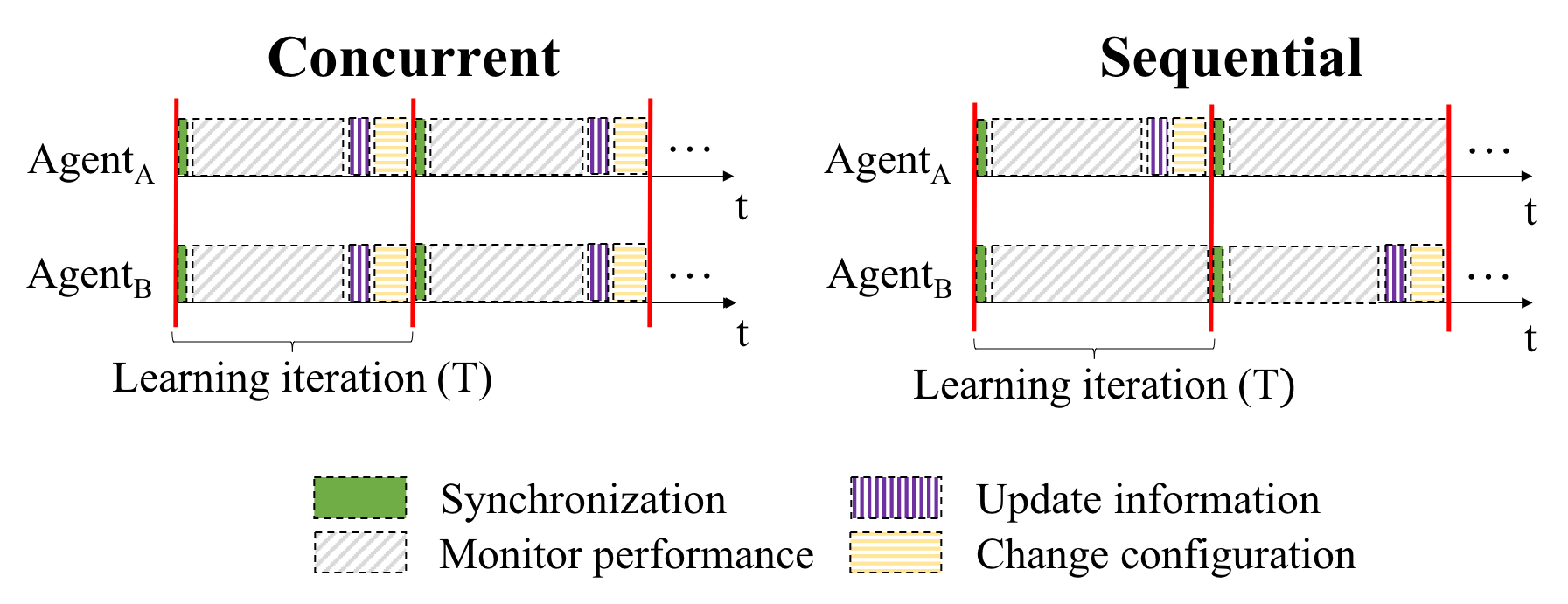, width=11cm}
		\caption{Concurrent and sequential procedures.}
		\label{fig:async_vs_sync}
	\end{figure}

	\textcolor{black}{We frame the decentralized learning procedure in two different ways, namely \textit{concurrent} and \textit{sequential}. Figure \ref{fig:async_vs_sync} illustrates the procedure followed by agents to carry out decentralized SR learning. As shown, in each iteration\footnote{\textcolor{black}{The time between iterations ($T$) must be large enough to provide an accurate estimation of the throughput experienced for a given action profile.}} there is a monitoring phase (shown in grey), where the current selected action is analyzed by each agent to quantify the hidden reward (which depends on the adversarial setting). Such a reward is the same for all the policies presented in this work, so that a fair comparison can be provided. After the monitoring phase is completed, agents update their knowledge (shown in purple) and choose a new action (shown in yellow). Note, as well, that both approaches rely on a synchronization phase (shown in green), which can be achieved  through message passing \cite{sanghavi2009message, yang2011message} and/or environment sensing.\footnote{\textcolor{black}{The IEEE 802.11k amendment, which is devoted to measurement reporting, may enable the environment sensing operation.}}} 
	
	\textcolor{black}{In the concurrent approach, agents (or WNs) make decisions simultaneously, thus leading to a more variable (and chaotic) environment. In practice the fully decentralized learning process will most probably not be synchronized but we leave the study of any possible effects of that desynchronization to future work. In contrast, \textcolor{black}{for the sequential approach, WNs need to wait for their turn in order to pick a new action. As a result, the performance of their last selected action (arm) is measured for several iterations (equal to the number of overlapping networks). In particular, during the update phase, a WN computes the reward of its last selected arm according to the throughput experienced in average. Accordingly, the performance of a given action is measured against different adversarial settings, since the environment changes gradually. Despite agents still learn selfishly, they can better assess how robust an action is against the joint actions profile, in comparison to the concurrent approach.}} 
	
	\subsection{Simulation Parameters}
	\label{section:simulation_parameters}
	According to \cite{bellalta2016ax}, which provides an overview of the IEEE 802.11ax-2019 standard, a typical high-density scenario for residential buildings contains $0.0033 \text{APs}/\text{m}^3$ (i.e., 100 APs in a $100 \times 20 \times 15$ m area). Accordingly, for simulation purposes, we define a map scenario with dimensions $10\times5\times10$ m, containing from 2 to 8 APs. In addition, for the first part of the simulations, we consider a setting containing 4 WNs that form a grid topology. In it, STAs are placed at the maximum possible distance from the other networks. Table \ref{tbl:simulation_parameters} details the parameters used.	
	\begin{table}[h!]
		\centering
		\resizebox{0.6\columnwidth}{!}{
			\begin{tabular}{|l|l|}
				\hline
				\textbf{Parameter}             & \textbf{Value}                      \\ \hline
				Map size (m)                    & $10\times5\times10$                  \\ \hline
				Number of coexistent WNs      & \{2, 4, 6, 8\} \\ \hline
				APs/STAs per WN                    & 1 / 1                                     \\ \hline
				Distance AP-STA (m)     & $\sqrt{2}$                         \\ \hline
				Number of \textcolor{black}{orthogonal} channels              & \textcolor{black}{3}                               \\ \hline
				Channel bandwidth (MHz)         & 20                                    \\ \hline
				Initial channel selection model & Uniformly distributed 
				\\ \hline
				\textcolor{black}{Transmit power values (dBm)}                & \textcolor{black}{\{-15, 0, 15, 30\}}                      \\ \hline
				$\text{PL}_0$ (dB)                   & 5
				\\ \hline
				$\text{G}_s$ (dB)                   & Normally distributed with mean 9.5 
				\\ \hline
				$\text{G}_o$ (dB)                  & Uniformly distributed with mean 30
				\\ \hline
				$d_{\rm obs}$  (meters between two obstacles)                 & 5                        
				\\ \hline
				Noise level (dBm)               & -100                                  \\ \hline
				Traffic model                   & Full buffer (downlink)             \\ \hline          
				Number of learning iterations                  & 10,000             \\ \hline   
		\end{tabular}}
		\caption{Simulation parameters}
		\label{tbl:simulation_parameters}
	\end{table}
	
	\section{Performance Evaluation}
	\label{section:performance_evaluation}
	
	In this Section, we evaluate the performance of each action-selection strategy presented in Section \ref{section:mabs} when applied to the decentralized SR problem in WNs.\footnote{The source code used in this work is open \cite{fwilhelmi2017code}, encouraging sharing of \textcolor{black}{knowledge} with potential contributors under the GNU General Public License v3.0.} For that purpose, we first evaluate in Section \ref{section:toy_grid_scenario} the $\varepsilon$-greedy, EXP3, UCB and Thompson sampling policies in a fixed adversarial environment. \textcolor{black}{This allows us to provide insights on the decentralized learning problem in a competitive scenario. Accordingly, we are able to analyze in detail the effect of applying each learning policy on the network's performance.} Without loss of generality, we consider a symmetric configuration and analyze the competition effects when WNs have the same opportunities for accessing the channel. Finally, Section \ref{section:random} provides a performance comparison of the aforementioned scenarios with different densities and with randomly located WNs.	
	\subsection{Toy Grid Scenario}		
	\label{section:toy_grid_scenario}	
	
\begin{figure}[h!]
		\centering								
		\epsfig{file=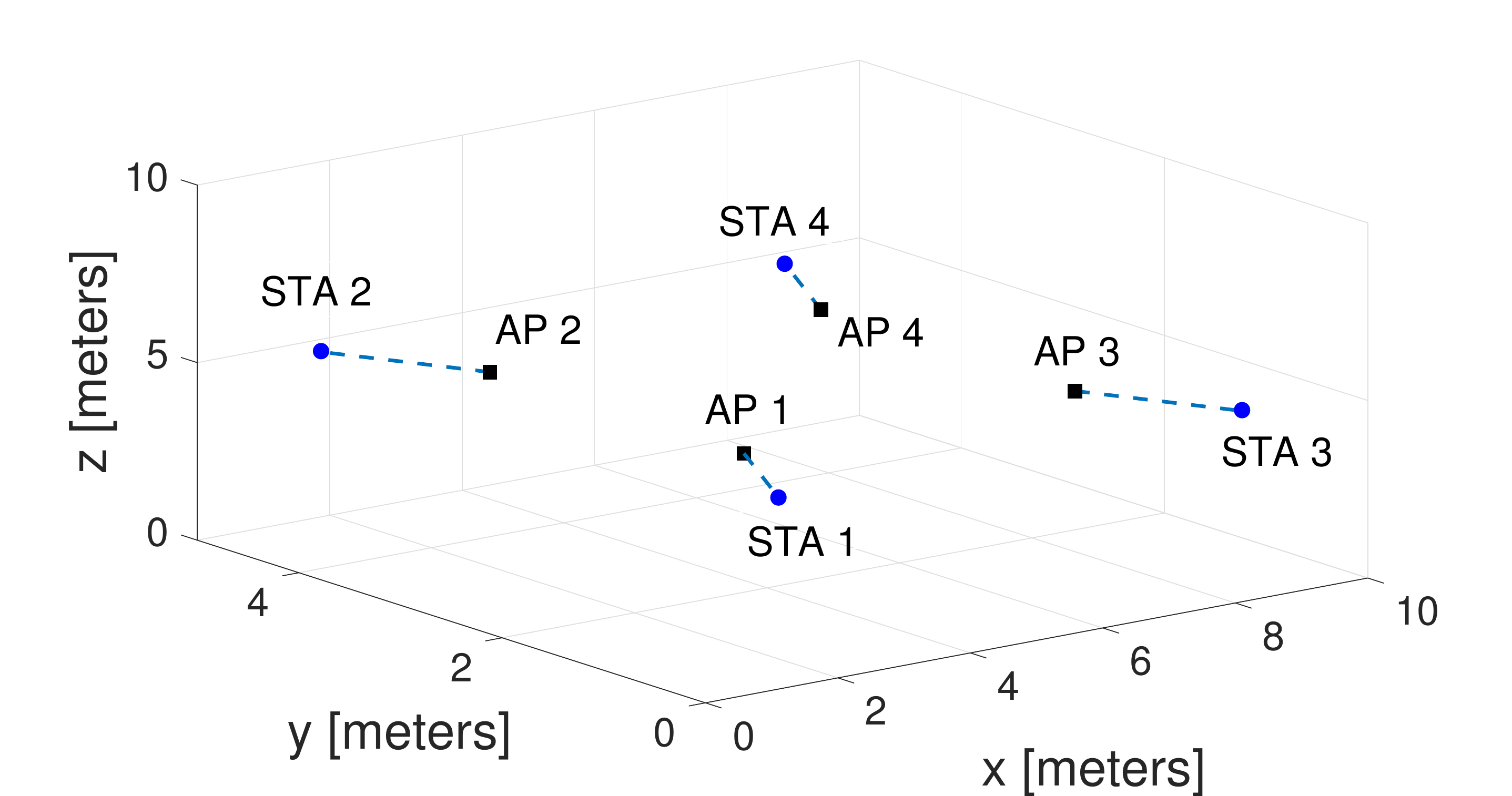, width=8.5cm}
		\caption{Grid scenario containing 4 WNs, each one composed by an AP and a STA.}
		\label{fig:scenario}
	\end{figure}

	The toy grid scenario contains 4 WNs and is illustrated in Figure \ref{fig:scenario}. This scenario has the particularity of being symmetric, so that adversarial WNs have the same opportunities to compete for the channel resources. \textcolor{black}{The optimal solution in terms of proportional fairness\footnote{\textcolor{black}{The proportional fairness (PF) result accomplishes that the logarithmic sum of each individual throughput is maximized: $\max \sum_{i \in \text{WN}} \log(\Gamma_i)$.}} is achieved when channel reuse is maximized and WNs sharing the channel moderate their transmit power. The PF solution provides an aggregate performance of $440.83$ Mbps (i.e., $106.212$ Mbps per WN on average). The optimal solution is computed by brute force (i.e., trying all the combinations), and it is used as a baseline.}
	
	\textcolor{black}{\subsubsection{Configuration of the Learning Parameters}}
	
	\textcolor{black}{Before comparing the performance of each algorithm, we first analyze the effect of modifying each one's internal parameters. Since the versions of UCB and Thompson sampling analyzed in this work are parameter-less, in this section we focus only on $\varepsilon$-greedy and EXP3 methods.}
	
	\textcolor{black}{Firstly, $\varepsilon$-greedy allows to regulate the explicit exploration rate at which the agent operates, which is referred to as $\varepsilon$. In this paper, $\varepsilon$ is dynamically adjusted as $\varepsilon_t = \frac{\varepsilon_0}{\sqrt{t}}$, with the aim of exploring more efficiently. Accordingly, we study the impact of modifying the initial exploration coefficient in the experienced performance by a WN. Secondly, when it comes to EXP3, we find two parameters, namely $\eta$ and $\gamma$. While $\eta$ controls how fast old beliefs are replaced by newer ones, $\gamma$ regulates explicit exploration by tuning the importance of weights in the action-selection procedure. Setting $\gamma = 1$ results in completely neglecting weights (actions have the same probability to be chosen). On the other side, by setting $\gamma = 0$, the effect of weights are at its highest importance. Thus, in order to clearly analyze the effects of the EXP3 weights, which directly depend on $\eta$, we fix $\gamma$ to 0. As we did for $\varepsilon$-greedy, we analyze the impact of modifying the parameter $\eta_0$ in EXP3 on the WN's performance.}
	
\begin{figure}[h!]
		\centering
		\epsfig{file=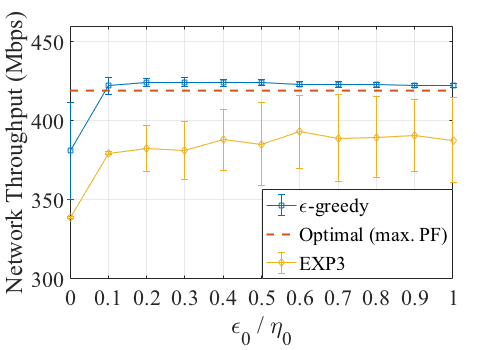, width=7cm}
		\label{fig:tuning_parameters}
		\caption{\textcolor{black}{Average network throughput and standard deviation obtained for each $\varepsilon_0$ and $\eta_0$ value in $\varepsilon$-greedy and EXP3, respectively. Results are from 100 simulations lasting 10,000 iterations each. The proportional fair solution is also shown (red dashed line).}}
	\end{figure}

	\textcolor{black}{Figure \ref{fig:tuning_parameters} shows the aggregate throughput obtained in the grid scenario when applying both $\varepsilon$-greedy and EXP3 during 10,000 iterations, and for each $\varepsilon_0$ and $\eta_0$ values, respectively. The results are presented for values $\varepsilon_0$ and $\eta_0$ between 0 and 1 in 0.1 steps. The average and standard deviation of the throughput from 100 simulation runs are also shown, and compared with the proportional fair solution.}
	
	\textcolor{black}{As shown, the aggregate throughput obtained on average is quite similar for all $\varepsilon_0$ and $\eta_0$ values, except for the complete random case where no exploration is done (i.e., when $\varepsilon_0$ and $\eta_0$ are equal to 0). For $\varepsilon$-greedy, the lower the $\varepsilon_0$ parameter, the less exploration is performed. Consequently, for low $\varepsilon_0$, the average throughput is highly dependent on how good/bad were the actions taken at the beginning of the learning process, which results in a higher standard deviation as $\varepsilon_0$ goes to 0. As for EXP3, the lower $\eta_0$, the more slowly weights are updated. For $\eta_0 = 0$, weights are never updated, so that arms have always the same probability to be chosen. To conclude, we choose $\varepsilon_0 = 1$ and $\eta_0 = 0.1$, respectively, for the rest of simulations, which provide the highest ratio between the aggregate throughput and the variability among different runs.}
	
	\textcolor{black}{\subsubsection{Performance of the MAB-based Policies}}
	
\begin{figure}[h!]
		\centering
		\begin{subfigure}[b]{0.45\textwidth}
			\includegraphics[width=\textwidth]{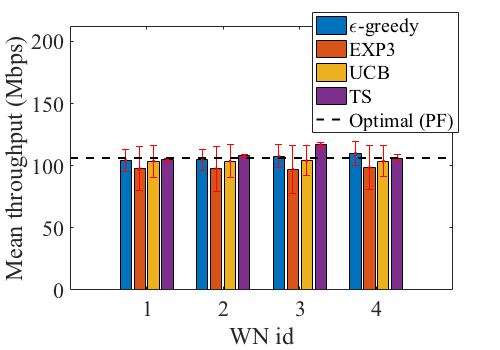}
			\caption{Mean throughput}
			\label{fig:results_part_1_variability_default}
		\end{subfigure}
		\begin{subfigure}[b]{0.45\textwidth}
			\includegraphics[width=\textwidth]{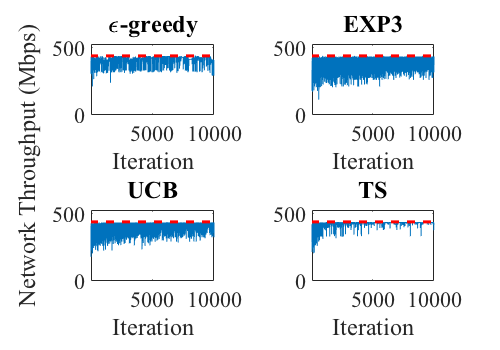}
			\caption{Temporal network throughput}
			\label{fig:results_part_1_agg_variability_default}
		\end{subfigure}
		\caption{\textcolor{black}{Mean throughput achieved per WN, for each action-selection strategy (the standard deviation is shown in red). The black dashed line indicates the PF result.}}
		\label{fig:results_part_1_mean_tpt}
	\end{figure}

	\textcolor{black}{Once we established the initial parameters to be used by both $\varepsilon$-greedy and EXP3, we now compare the performance of all the studied action-selection strategies when applied to decentralized WNs. First, we focus on the average throughput achieved by each WN in the toy grid scenario, for each of the methods (Figure \ref{fig:results_part_1_variability_default}). As shown, the proportional fair solution is almost achieved by all the learning methods. However, Thompson sampling is shown to be much more stable than the other mechanisms, since its variability in the aggregate throughput is much lower (depicted in Figure \ref{fig:results_part_1_agg_variability_default}).}
	
\begin{figure}[t!]
		\centering
		\begin{subfigure}[b]{0.43\textwidth}
			\includegraphics[width=\textwidth]{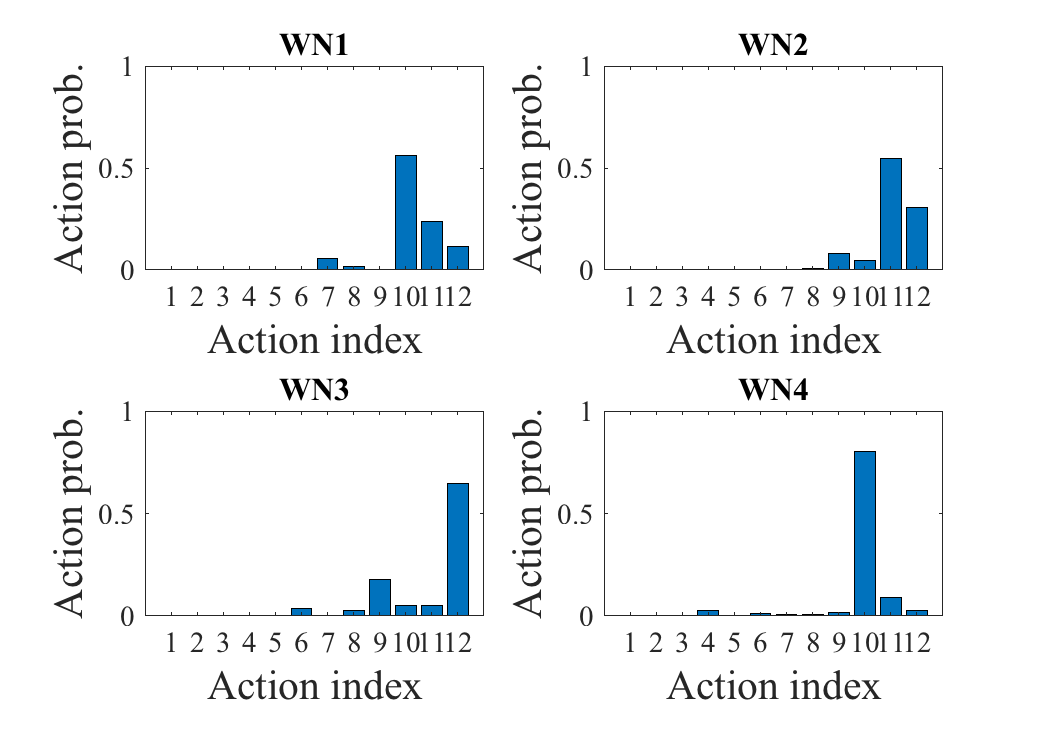}
			\caption{$\varepsilon$-greedy ($\varepsilon_0  = 0.1$)}
			\label{fig:actions_probability_EG}
		\end{subfigure}
		\begin{subfigure}[b]{0.43\textwidth}
			\includegraphics[width=\textwidth]{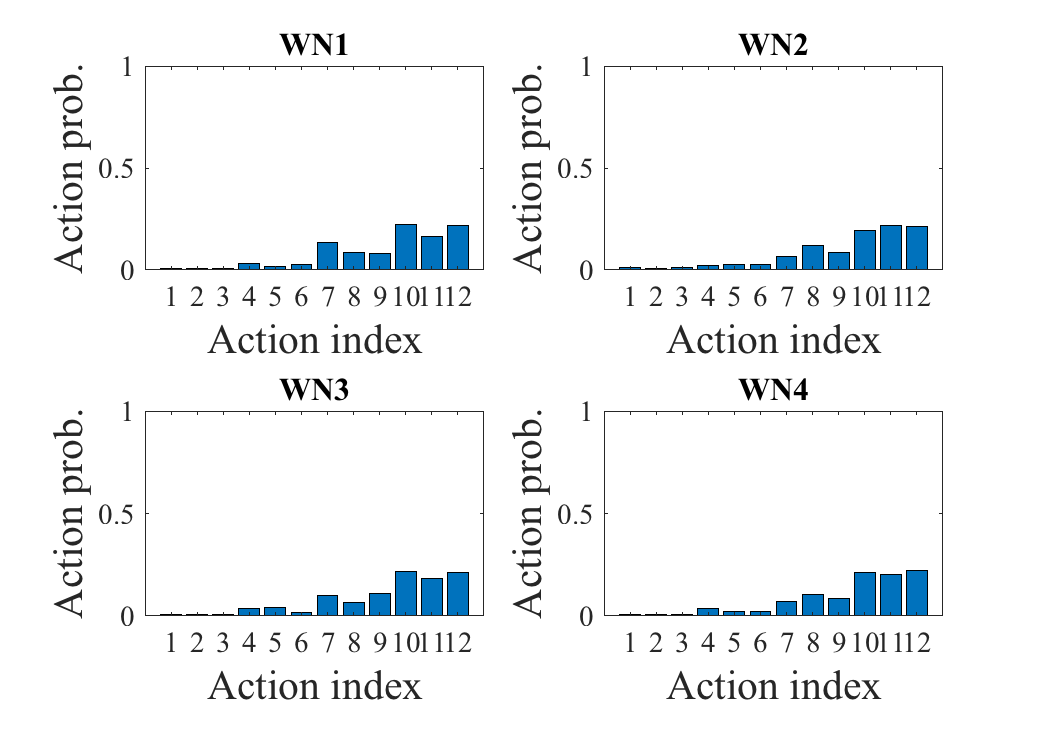}
			\caption{EXP3 ($\eta_0 = 0.1$)}
			\label{fig:actions_probability_EXP3}
		\end{subfigure}
		\begin{subfigure}[b]{0.43\textwidth}
			\includegraphics[width=\textwidth]{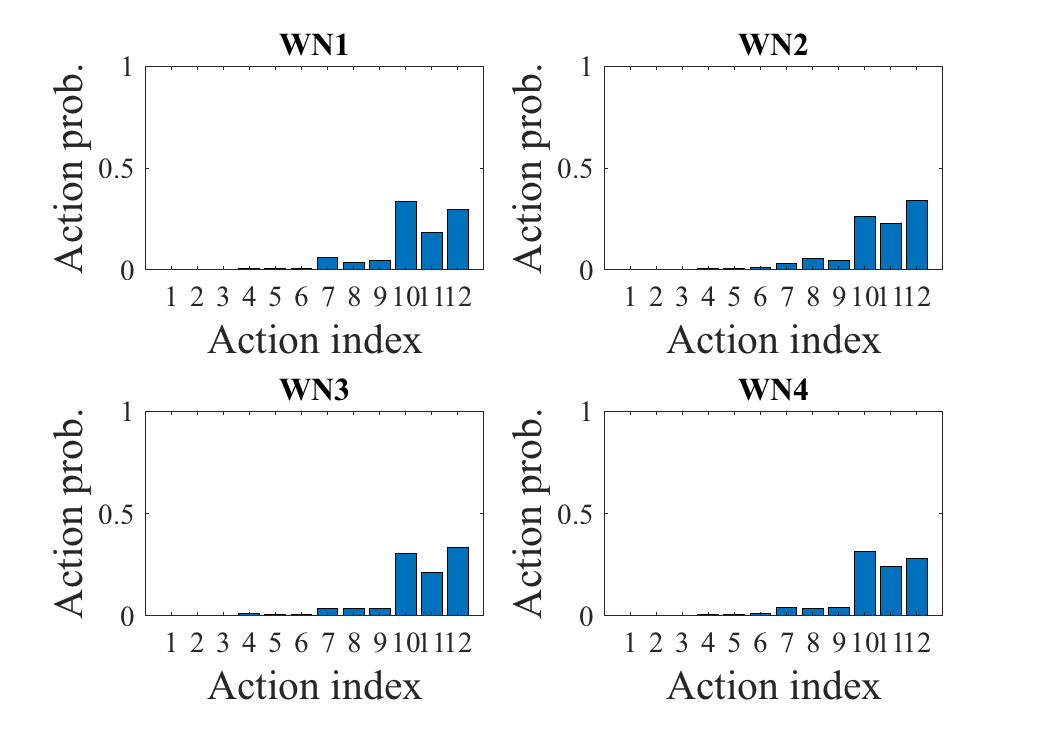}
			\caption{UCB}
			\label{fig:actions_probability_UCB}
		\end{subfigure}
		\begin{subfigure}[b]{0.43\textwidth}
			\includegraphics[width=\textwidth]{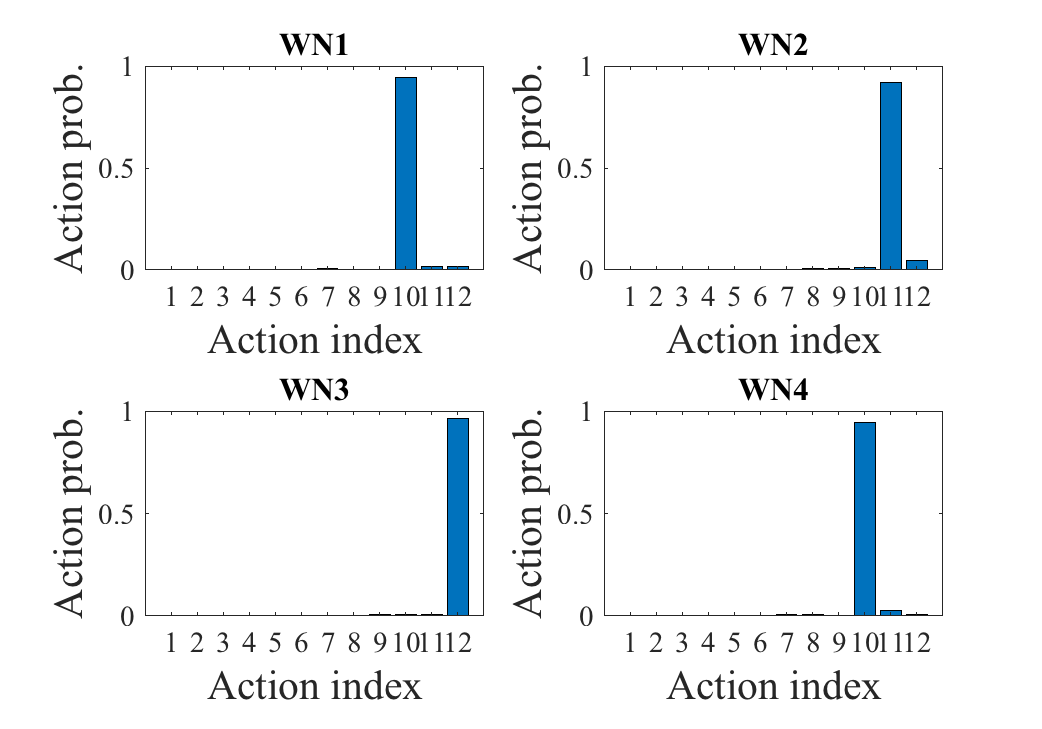}
			\caption{TS}
			\label{fig:actions_probability_TS}
		\end{subfigure}
		\caption{\textcolor{black}{Probability of selecting each given action for a simulation of 10,000 iterations.}}
		\label{fig:actions_probabilities}
	\end{figure}

	\textcolor{black}{In order to dig deeper into the behavior of agents for each policy, Figure \ref{fig:actions_probabilities} shows the probability of each WN to choose each action. Regarding $\varepsilon$-greedy, EXP3 and UCB, a large set of actions is chosen with similar probabilities. Note that there are only three frequency channels, so that two WNs need to share one of them, thus leading to a lower performance with respect to the other two. Therefore, WNs are constantly changing their channel and experiencing intermittent good/poor performance. Thus, the degree of exploration is kept very high, resulting in high temporal variability. In contrast, Thompson sampling shows a clearer preference for selecting a single action, which allows reducing the aforementioned variability.}
	
	\textcolor{black}{\subsubsection{Learning Sequentially}}
	\label{section:proposed_method}
	
\begin{figure}[t!]
		\centering	
		\begin{subfigure}[b]{0.44\textwidth}
			\includegraphics[width=\textwidth]{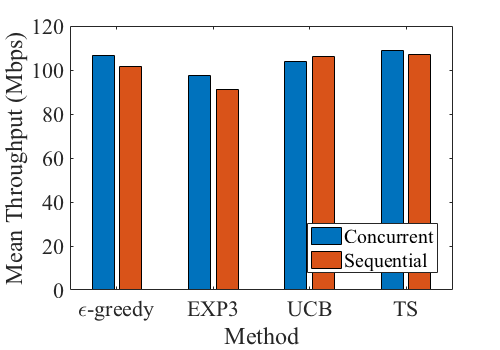}
			\caption{Average throughput}
			\label{fig:results_part_1_async_vs_sync_meant_tpt}
		\end{subfigure}
		\begin{subfigure}[b]{0.44\textwidth}
			\includegraphics[width=\textwidth]{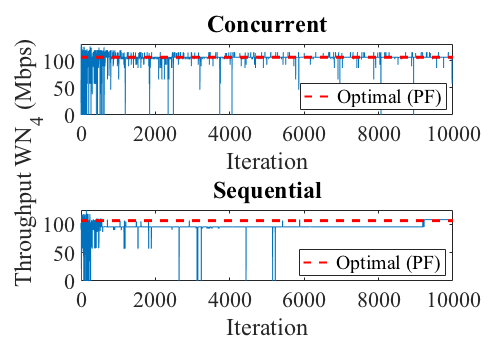}
			\caption{Temporal variability in $\text{WN}_4$}
			\label{fig:results_part_1_async_vs_sync_variability}
		\end{subfigure}
		\caption{\textcolor{black}{Concurrent vs sequential approaches performance. (a) Mean average throughput achieved for each learning procedure. (b) Temporal variability experienced by $\text{WN}_4$ for the Thompson sampling action-selection strategy and for each learning procedure.}}
		\label{fig:results_part_1_async_vs_sync}
	\end{figure}

	\textcolor{black}{In order to alleviate the strong throughput variability experienced when applying decentralized learning, we now focus on the sequential approach introduced in Section \ref{section:learning_procedure}. Now, only one WN is able to select an action at a time. With that, we aim to reduce the adversarial effect on the estimated rewards. Therefore, by having a more stable environment (not all the WNs learn simultaneously), the actual reward of a given selected action can be estimated more accurately. Figure \ref{fig:results_part_1_async_vs_sync} shows the differences between learning through concurrent and sequential mechanisms. Firstly, the throughput experienced on average along the entire simulation is depicted in Figure \ref{fig:results_part_1_async_vs_sync_meant_tpt}. Secondly, without loss of generality, Figure \ref{fig:results_part_1_async_vs_sync_variability} shows the temporal variability experienced by $\text{WN}_4$ when applying Thompson sampling. Note that showing the performance of a single WN is representative enough for the entire set of WNs (the scenario is symmetric), and allows us to analyze in detail the behavior of the algorithms.}
	
	\textcolor{black}{On the one hand, a lower throughput is experienced on average when learning in a sequential way, but the differences are very small. In such a situation, WNs spend more time observing sub-optimal actions, since they need to wait for their turn. Note, as well, that the time between iterations ($T$) depends on the implementation. In this particular case, we assume that $T$ is the same for both sequential and concurrent approaches.}
	
	\textcolor{black}{On the other hand, the temporal variability shown by the sequential approach is much lower than for the concurrent one (Figure \ref{fig:results_part_1_async_vs_sync_variability}). The high temporal variability may negatively impact on the user's experience and the operation of upper layer protocols (e.g., TCP) may be severely affected. Notice that a similar effect is achieved for the rest of algorithms.}
	
	\subsubsection{Learning in a Dynamic Environment}
	\label{section:dynamic_environment}
	
	\textcolor{black}{Finally, we show the performance of the proposed learning mechanisms in a dynamic scenario. For that, we propose the following situation. Firstly, $\text{WN}_1$ and $\text{WN}_2$ are active for the whole simulation. Secondly, $\text{WN}_3$ turns on at iteration 2,500, when $\text{WN}_1$ and $\text{WN}_2$ are supposed to have acquired enough knowledge to maximize SR. Finally, $\text{WN}_4$ turns on at iteration 5,000, similarly than for $\text{WN}_3$.}
	
\begin{figure}[h!]
		\centering
		\begin{subfigure}[b]{0.45\textwidth}
			\includegraphics[width=\textwidth]{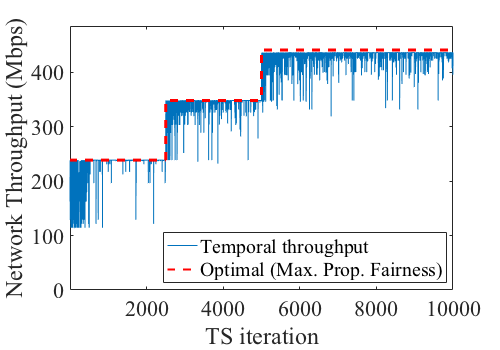}
			\caption{Concurrent approach}
			\label{fig:temporal_aggregate_tpt_dynamic_scenario_async_TS}
		\end{subfigure}
		\begin{subfigure}[b]{0.45\textwidth}
			\includegraphics[width=\textwidth]{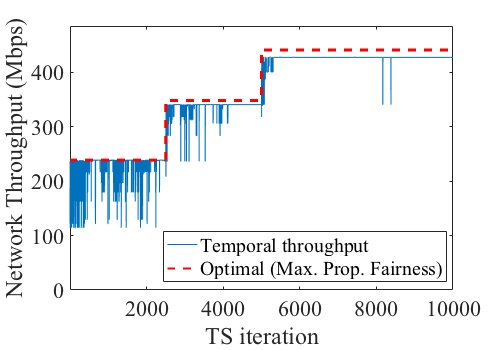}
			\caption{Sequential approach}
			\label{fig:temporal_aggregate_tpt_dynamic_scenario_ordered_TS}
		\end{subfigure}
		\caption{\textcolor{black}{Temporal aggregate throughput experienced for a 10,000-iteration Thompson sampling simulation.}}
		\label{fig:dynamic_enviroment}
	\end{figure}

	\textcolor{black}{Through this simulation, we aim to show how each learning algorithm adapts to changes in the environment, which highly impact on the rewards distributions. Figure \ref{fig:dynamic_enviroment} shows the temporal aggregate throughput achieved by each action-selection strategy. As done in Subsection \ref{section:proposed_method}, we only plot the results of the best-performing algorithm, i.e., Thompson sampling, both for the concurrent and the sequential procedures.}
	
	\textcolor{black}{As shown, WNs are able to adapt to the changes in the environment. In particular, for the concurrent case (see Figure \ref{fig:temporal_aggregate_tpt_dynamic_scenario_async_TS}), changes are harder to be captured as the network size increases. In contrast, learning in an ordered way (see Figure \ref{fig:temporal_aggregate_tpt_dynamic_scenario_ordered_TS}) allows reducing the temporal variability, even if new WNs turn on. However, there is a little loss in the aggregate performance with respect to the concurrent approach. The difference between the maximum network performance is mostly provoked by the reduced exploration shown by the sequential approach.}
	\subsection{Random Scenarios}
	\label{section:random}
	
\begin{figure}[h!]
		\centering
		\includegraphics[width=0.8\textwidth]{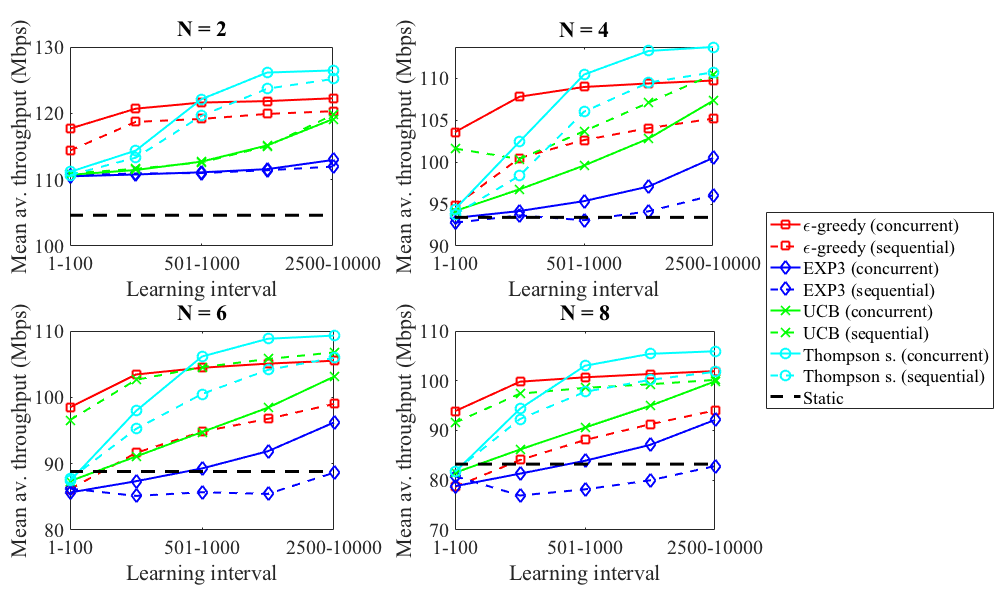}
		\caption{\textcolor{black}{Average throughput experienced in each learning interval for each action-selection strategy. Results from 100 repetitions are considered for each different number of overlapping WNs (N = \{2, 4, 6, 8\}). The black dashed line indicates the default IEEE 802.11 performance (static situation).}}
		\label{fig:random_scenarios_results}
	\end{figure}

\begin{figure}[h!]
		\centering
		\includegraphics[width=0.77\textwidth]{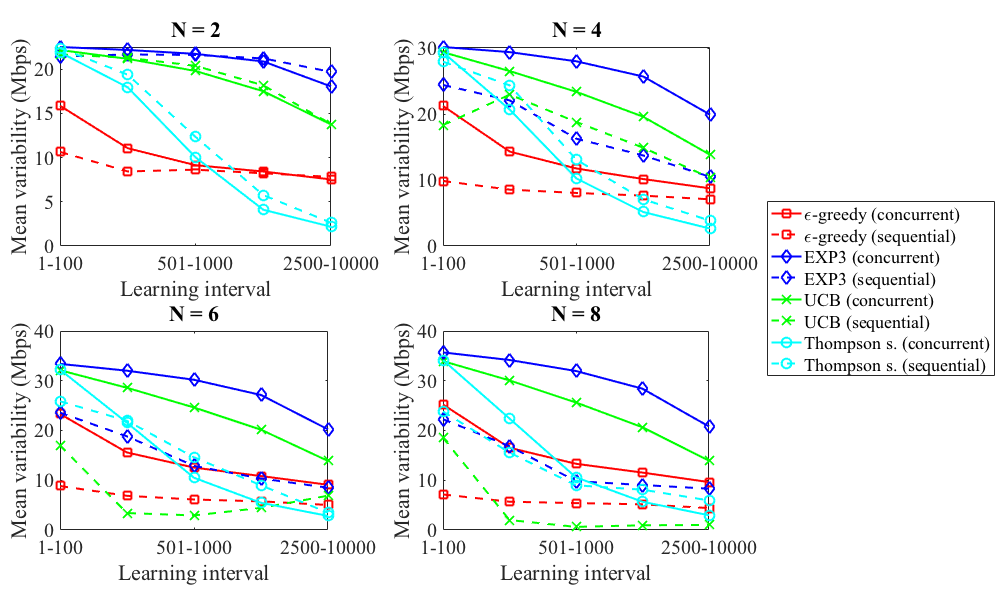}
		\caption{\textcolor{black}{Average variability experienced in each learning interval for each action-selection strategy. Results from 100 repetitions are considered for each different number of overlapping WNs (N = \{2, 4, 6, 8\}).}}
		\label{fig:random_scenarios_results_variability}
	\end{figure}

	We now evaluate whether the previous conclusions generalize to random scenarios with an arbitrary number of WNs. To this aim, we use the same $10\times5\times 10$ m scenario and randomly allocate N = \{2, 4, 6, 8\} WNs. \textcolor{black}{Figures \ref{fig:random_scenarios_results} and \ref{fig:random_scenarios_results_variability} show the mean throughput and variability experienced for each learning strategy, and for each number of coexistent WNs, respectively. The variability is measured as the standard deviation that a given WN experiences along an entire simulation. We consider the average results of 100 different random scenarios for each number of networks. In particular, we are interested on analyzing the gains achieved by each algorithm, even if convergence cannot be provided due to the competition between networks. For that, we display the average performance for the following learning intervals: [1-100, 101-500, 501-1000, 1001-2500, 2501-10000]. Note that the first intervals represent few iterations. This allows us to observe the performance achieved during the transitory phase in more detail. In addition, the performance achieved in a static situation (i.e., when no learning is performed) is shown in Figure \ref{fig:random_scenarios_results}. With that, we aim to compare the gains obtained by each learning strategy with respect to the current IEEE 802.11 operation in unplanned and chaotic deployments.}
	
	\textcolor{black}{First of all, let us focus on the throughput improvements achieved with respect to the static situation. As shown in Figure \ref{fig:random_scenarios_results}, each learning strategy easily outperforms the static scenario for low densities (i.e., 2 and 4 overlapping WNs). However, as density increases, improving the average throughput becomes more challenging. This is clearly evidenced for N = \{6, 8\} WNs, where EXP3 performs worse than the static situation.}
	
	\textcolor{black}{Secondly, we concentrate on the concurrent learning procedure. As shown in Figure \ref{fig:random_scenarios_results}, Thompson sampling outperforms the other action-selection strategies for all the scenarios, provided that enough exploration is done (up to 500 iterations). On the other hand, $\varepsilon$-greedy allows to increase the average performance very quickly, but its growth stalls from iteration 200. Note that $\varepsilon$-greedy is based on the absolute throughput value, thus preventing to find a collaborative behavior in which the scarce resources are optimally shared. Finally, EXP3 and UCB are shown to improve the average throughput linearly, but offering poor performance.}
	
	\textcolor{black}{When it comes to the sequential approach, we find the following:}
	\begin{itemize}
		\item \textcolor{black}{On the one hand, the average throughput is reduced in almost all the cases in comparison with the concurrent approach (see Figure \ref{fig:random_scenarios_results}). We find that this is due to the larger phases in which agents exploit sub-optimal actions. As previously pointed out, the time between iterations is considered to be the same for both sequential and concurrent learning approaches.}
		\item \textcolor{black}{On the other hand, the sequential procedure is shown to substantially improve the variability experienced by $\varepsilon$-greedy, EXP3 and UCB (see Figure \ref{fig:random_scenarios_results_variability}). The performance of the latter is particularly shown to be improved when the learning procedure is ordered. The sequential approach, therefore, allows UCB to produce more accurate estimates on the actions rewards. In contrast, learning sequentially does not improve the concurrent version of Thompson sampling in any performance metric. We attribute this suboptimal behavior to the way in which Thompson sampling performs estimates of the played actions, which depends on the number of times each one is selected. In particular, suboptimal actions can eventually provide good enough performance according to the adversarial setting. The same issue can lead to underestimate optimal actions, so that their actual potential is not observed. Since Thompson sampling bases its estimates on the number of times each action is selected, the aforementioned effect may lead to increase the exploitation on suboptimal actions.}
	\end{itemize}
	
	\section{Conclusions }
	\label{section:conclusions}
	In this paper, we provided an implementation of MABs to address the decentralized SR problem in dense WNs. Unlike previous literature, we have focused on a situation in which few resources are available, thus bringing out the competition issues raised from the adversarial setting. Our results show that decentralized learning allows improving SR in dense WN, so that collaborative results in symmetric scenarios, sometimes close to optimal proportional fairness, can be achieved. This result is achieved even though WNs act selfishly, aiming to maximize their own throughput. In addition, this behavior is observed for random scenarios, where the effects of asymmetries cannot be controlled. 
	These collaborative actions are, at times, accompanied by high temporal throughput variability, which can be understood as a consequence of the  rate at which networks change their configuration in response of the opponents behavior. A high temporal variability may provoke negative issues in a node's performance, as its effects may be propagated to higher layers of the protocol stack. For instance, a high throughput fluctuation may entail behavioral anomalies in protocols such as Transmission Control Protocol (TCP). 
	
	We have studied this trade-off between fair resource allocation and high temporal throughput variability in $\varepsilon$-greedy, EXP3, UCB and Thompson sampling action-selection strategies. Our results show that while this trade-off is hard to regulate via the learning parameters in $\varepsilon$-greedy and EXP3, UCB  and, especially, Thompson sampling are able to achieve fairness at a reduced temporal variability. We identify the root cause of this phenomena to the fact that both UCB and Thompson sampling consider the probability distribution of the rewards, and not only their magnitude.
	
	\textcolor{black}{Furthermore, for the sake of alleviating the temporal variability, we studied the effects of learning concurrent and sequentially. We have shown that learning in an ordered way is very effective to reduce the throughput variability for almost all the proposed learning strategies, even if WNs maintain a selfish behavior. By learning sequentially, more knowledge is attained on a given action, thus allowing to differentiate quickly between good and bad performing actions. \textcolor{black}{Apart from that, we found that Thompson sampling grants significantly better results than the other examined algorithms since it is able to capture meaningful information from chaotic environments.}}
	
	We left as future work to further study the MABs application to WNs through distributed (with message passing) and centralized (with complete information) approaches with shared reward. In particular, we would like to extend this work to enhance both throughput and stability by inferring the actions of the opponents and acting in consequence, as well as further investigating dynamic scenarios. Defining the resource allocation problem as an adversarial game is one possibility to do so. \textcolor{black}{In addition to this, the utilization of multiple antenna strategies (i.e., single and multi-user beamforming and interference cancellation) is expected to further improve the spectral efficiency in future WNs. Through these techniques, the SR problem can be relaxed in a similar way than using several non-overlapping frequency channels. However, its application would significantly increase the problem's complexity, and its analysis is also left as future work.}
	
	\section*{Acknowledgments}
	This work has been partially supported by the Spanish Ministry of Economy and Competitiveness under the Maria de Maeztu Units of Excellence Programme (MDM-2015-0502), by a Gift from CISCO University Research Program (CG\#890107) \& Silicon Valley Community Foundation, by the European Regional Development Fund under grant TEC2015-71303-R (MINECO/FEDER), and by the Catalan Government under grant SGR-2017-1188.
	
	\textcolor{black}{The authors would like to thank the anonymous reviewers. Their dedication and insightful comments were of great help to improve this paper. We would also like to show our gratitude to Andrés Burbano, who also contributed to improve the quality of this document through his thorough revision.
	}
	
	\newpage
	\bibliographystyle{unsrt}
	\bibliography{references}

\begin{thebibliography}{10}

\bibitem{akella2007self}
Aditya Akella, Glenn Judd, Srinivasan Seshan, and Peter Steenkiste.
\newblock Self-management in chaotic wireless deployments.
\newblock {\em Wireless Networks}, 13(6):737--755, 2007.

\bibitem{alawieh2009improving}
Basel Alawieh, Yongning Zhang, Chadi Assi, and Hussein Mouftah.
\newblock Improving spatial reuse in multihop wireless networks-a survey.
\newblock {\em IEEE Communications Surveys \& Tutorials}, 11(3), 2009.

\bibitem{miridakis2013survey}
Nikolaos~I Miridakis and Dimitrios~D Vergados.
\newblock A survey on the successive interference cancellation performance for
  single-antenna and multiple-antenna ofdm systems.
\newblock {\em IEEE Communications Surveys \& Tutorials}, 15(1):312--335, 2013.

\bibitem{dovelos2018breaking}
Konstantinos Dovelos and Boris Bellalta.
\newblock Breaking the interference barrier in dense wireless networks with
  interference alignment.
\newblock In {\em 2018 IEEE International Conference on Communications (ICC)},
  pages 1--6. IEEE, 2018.

\bibitem{littman1993distributed}
Michael Littman and Justin Boyan.
\newblock A distributed reinforcement learning scheme for network routing.
\newblock In {\em Proceedings of the international workshop on applications of
  neural networks to telecommunications}, pages 45--51. Psychology Press, 1993.

\bibitem{bojovic2011supervised}
Biljana Bojovic, Nicola Baldo, Jaume Nin-Guerrero, and Paolo Dini.
\newblock A supervised learning approach to cognitive access point selection.
\newblock In {\em GLOBECOM Workshops (GC Wkshps), 2011 IEEE}, pages 1100--1105.
  IEEE, 2011.

\bibitem{bojovic2012neural}
Biljana Bojovic, Nicola Baldo, and Paolo Dini.
\newblock A neural network based cognitive engine for ieee 802.11 {WLAN} access
  point selection.
\newblock In {\em Consumer Communications and Networking Conference (CCNC),
  2012 IEEE}, pages 864--868. IEEE, 2012.

\bibitem{combes2014optimal}
Richard Combes, Alexandre Proutiere, Donggyu Yun, Jungseul Ok, and Yung Yi.
\newblock Optimal rate sampling in 802.11 systems.
\newblock In {\em INFOCOM, 2014 Proceedings IEEE}, pages 2760--2767. IEEE,
  2014.

\bibitem{miozzo2015distributed}
Marco Miozzo, Lorenza Giupponi, Michele Rossi, and Paolo Dini.
\newblock Distributed q-learning for energy harvesting heterogeneous networks.
\newblock In {\em Communication Workshop (ICCW), 2015 IEEE International
  Conference on}, pages 2006--2011. IEEE, 2015.

\bibitem{BCB12}
S{\'e}bastien Bubeck and Nicol\'o Cesa-Bianchi.
\newblock Regret analysis of stochastic and nonstochastic multi-armed bandit
  problems.
\newblock {\em Foundations and Trends in Machine Learning}, 5(1):1--122, 2012.

\bibitem{wilhelmi2017implications}
Francesc Wilhelmi, Boris Bellalta, Cristina Cano, and Anders Jonsson.
\newblock Implications of decentralized q-learning resource allocation in
  wireless networks.
\newblock In {\em Personal, Indoor, and Mobile Radio Communications (PIMRC),
  2017 IEEE 28th Annual International Symposium on}, pages 1--5. IEEE, 2017.

\bibitem{argyriou2010collision}
Antonios Argyriou and Ashish Pandharipande.
\newblock Collision recovery in distributed wireless networks with
  opportunistic cooperation.
\newblock {\em IEEE Communications Letters}, 14(4), 2010.

\bibitem{babich2015design}
Fulvio Babich, Massimiliano Comisso, Alessandro Crismani, and Aljo{\v{s}}a
  Dorni.
\newblock On the design of mac protocols for multi-packet communication in ieee
  802.11 heterogeneous networks using adaptive antenna arrays.
\newblock {\em IEEE Transactions on Mobile Computing}, 14(11):2332--2348, 2015.

\bibitem{riihijarvi2005frequency}
Janne Riihijarvi, Marina Petrova, and Petri Mahonen.
\newblock Frequency allocation for {WLAN}s using graph colouring techniques.
\newblock In {\em Wireless On-demand Network Systems and Services, 2005. WONS
  2005. Second Annual Conference on}, pages 216--222. IEEE, 2005.

\bibitem{mishra2005weighted}
Arunesh Mishra, Suman Banerjee, and William Arbaugh.
\newblock Weighted coloring based channel assignment for {WLAN}s.
\newblock {\em ACM SIGMOBILE Mobile Computing and Communications Review},
  9(3):19--31, 2005.

\bibitem{akl2007dynamic}
Robert Akl and Anurag Arepally.
\newblock Dynamic channel assignment in {IEEE} 802.11 networks.
\newblock In {\em Portable Information Devices, 2007. PORTABLE07. IEEE
  International Conference on}, pages 1--5. IEEE, 2007.

\bibitem{chen2007improved}
Jeremy~K Chen, Gustavo De~Veciana, and Theodore~S Rappaport.
\newblock Improved measurement-based frequency allocation algorithms for
  wireless networks.
\newblock In {\em Global Telecommunications Conference, 2007. GLOBECOM'07.
  IEEE}, pages 4790--4795. IEEE, 2007.

\bibitem{yue2011cacao}
Xiaonan Yue, Chi-Fai Wong, and S-H~Gary Chan.
\newblock Cacao: Distributed client-assisted channel assignment optimization
  for uncoordinated wlans.
\newblock {\em IEEE Transactions on Parallel and Distributed Systems},
  22(9):1433--1440, 2011.

\bibitem{elbatt2000power}
Tamer~A ElBatt, Srikanth~V Krishnamurthy, Dennis Connors, and Son Dao.
\newblock Power management for throughput enhancement in wireless ad-hoc
  networks.
\newblock In {\em Communications, 2000. ICC 2000. 2000 IEEE International
  Conference on}, volume~3, pages 1506--1513. IEEE, 2000.

\bibitem{tang2014joint}
Suhua Tang, Hiroyuki Yomo, Akio Hasegawa, Tatsuo Shibata, and Masayoshi Ohashi.
\newblock Joint transmit power control and rate adaptation for wireless {LAN}s.
\newblock {\em Wireless personal communications}, 74(2):469--486, 2014.

\bibitem{gandarillas2014dynamic}
Carlos Gandarillas, Carlos Mart{\'\i}n-Enge{\~n}os, H{\'e}ctor~L{\'o}pez Pombo,
  and Antonio~G Marques.
\newblock Dynamic transmit-power control for {WiFi} access points based on
  wireless link occupancy.
\newblock In {\em Wireless Communications and Networking Conference (WCNC),
  2014 IEEE}, pages 1093--1098. IEEE, 2014.

\bibitem{coucheney2015multi}
Pierre Coucheney, Kinda Khawam, and Johanne Cohen.
\newblock Multi-armed bandit for distributed inter-cell interference
  coordination.
\newblock In {\em ICC}, pages 3323--3328, 2015.

\bibitem{maghsudi2015joint}
Setareh Maghsudi and S{\l}awomir Sta{\'n}czak.
\newblock Joint channel selection and power control in infrastructureless
  wireless networks: A multiplayer multiarmed bandit framework.
\newblock {\em IEEE Transactions on Vehicular Technology}, 64(10):4565--4578,
  2015.

\bibitem{maghsudi2015channel}
Setareh Maghsudi and S{\l}awomir Sta{\'n}czak.
\newblock Channel selection for network-assisted {D2D} communication via
  no-regret bandit learning with calibrated forecasting.
\newblock {\em IEEE Transactions on Wireless Communications}, 14(3):1309--1322,
  2015.

\bibitem{thompson1933likelihood}
William~R Thompson.
\newblock On the likelihood that one unknown probability exceeds another in
  view of the evidence of two samples.
\newblock {\em Biometrika}, 25(3/4):285--294, 1933.

\bibitem{lai1985asymptotically}
Tze~Leung Lai and Herbert Robbins.
\newblock Asymptotically efficient adaptive allocation rules.
\newblock {\em Advances in applied mathematics}, 6(1):4--22, 1985.

\bibitem{auer2002finite}
Peter Auer, Nicol\'o Cesa-Bianchi, and Paul Fischer.
\newblock Finite-time analysis of the multiarmed bandit problem.
\newblock {\em Machine learning}, 47(2-3):235--256, 2002.

\bibitem{auer1995gambling}
Peter Auer, Nicol\'o Cesa-Bianchi, Yoav Freund, and Robert~E Schapire.
\newblock Gambling in a rigged casino: The adversarial multi-armed bandit
  problem.
\newblock In {\em Foundations of Computer Science, 1995. Proceedings., 36th
  Annual Symposium on}, pages 322--331. IEEE, 1995.

\bibitem{auer2002nonstochastic}
Peter Auer, Nicol\'o Cesa-Bianchi, Yoav Freund, and Robert~E Schapire.
\newblock The nonstochastic multiarmed bandit problem.
\newblock {\em SIAM journal on computing}, 32(1):48--77, 2002.

\bibitem{whittle1988restless}
Peter Whittle.
\newblock Restless bandits: Activity allocation in a changing world.
\newblock {\em Journal of applied probability}, 25(A):287--298, 1988.

\bibitem{LCLS10}
Lihong Li, Wei Chu, John Langford, and Robert~E Schapire.
\newblock A contextual-bandit approach to personalized news article
  recommendation.
\newblock In {\em Proceedings of the 19th international conference on World
  wide web}, pages 661--670. ACM, 2010.

\bibitem{abe2003reinforcement}
Naoki Abe, Alan~W Biermann, and Philip~M Long.
\newblock Reinforcement learning with immediate rewards and linear hypotheses.
\newblock {\em Algorithmica}, 37(4):263--293, 2003.

\bibitem{APS11}
Yasin Abbasi-Yadkori, D{\'a}vid P{\'a}l, and Csaba Szepesv{\'a}ri.
\newblock Improved algorithms for linear stochastic bandits.
\newblock In {\em Advances in Neural Information Processing Systems}, pages
  2312--2320, 2011.

\bibitem{sutton1998reinforcement}
Richard~S Sutton and Andrew~G Barto.
\newblock {\em Reinforcement learning: An introduction}, volume~1.
\newblock MIT press Cambridge, 1998.

\bibitem{Agr95}
Rajeev Agrawal.
\newblock Sample mean based index policies by o (log n) regret for the
  multi-armed bandit problem.
\newblock {\em Advances in Applied Probability}, 27(4):1054--1078, 1995.

\bibitem{BuKa96}
Apostolos~N Burnetas and Michael~N Katehakis.
\newblock Optimal adaptive policies for sequential allocation problems.
\newblock {\em Advances in Applied Mathematics}, 17(2):122--142, 1996.

\bibitem{LW94}
Nick Littlestone and Manfred~K Warmuth.
\newblock The weighted majority algorithm.
\newblock {\em Information and computation}, 108(2):212--261, 1994.

\bibitem{FS97}
Yoav Freund and Robert~E Schapire.
\newblock A decision-theoretic generalization of on-line learning and an
  application to boosting.
\newblock {\em Journal of computer and system sciences}, 55(1):119--139, 1997.

\bibitem{CL11}
Olivier Chapelle and Lihong Li.
\newblock An empirical evaluation of thompson sampling.
\newblock In J.~Shawe-Taylor, R.~S. Zemel, P.~L. Bartlett, F.~Pereira, and
  K.~Q. Weinberger, editors, {\em Advances in Neural Information Processing
  Systems 24}, pages 2249--2257. Curran Associates, Inc., 2011.

\bibitem{AG12}
Shipra Agrawal and Navin Goyal.
\newblock Analysis of thompson sampling for the multi-armed bandit problem.
\newblock In {\em Conference on Learning Theory}, pages 39--1, 2012.

\bibitem{KKM12}
Emilie Kaufmann, Nathaniel Korda, and R{\'e}mi Munos.
\newblock Thompson sampling: An asymptotically optimal finite time analysis.
\newblock In {\em Algorithmic Learning Theory}, pages 199--213. Springer, 2012.

\bibitem{KKM13}
Nathaniel Korda, Emilie Kaufmann, and Remi Munos.
\newblock Thompson sampling for 1-dimensional exponential family bandits.
\newblock In C.~J.~C. Burges, L.~Bottou, M.~Welling, Z.~Ghahramani, and K.~Q.
  Weinberger, editors, {\em Advances in Neural Information Processing Systems
  26}, pages 1448--1456. Curran Associates, Inc., 2013.

\bibitem{agrawal2013further}
Shipra Agrawal and Navin Goyal.
\newblock Further optimal regret bounds for thompson sampling.
\newblock In {\em Artificial Intelligence and Statistics}, pages 99--107, 2013.

\bibitem{sanghavi2009message}
Sujay Sanghavi, Devavrat Shah, and Alan~S Willsky.
\newblock Message passing for maximum weight independent set.
\newblock {\em IEEE Transactions on Information Theory}, 55(11):4822--4834,
  2009.

\bibitem{yang2011message}
Kai Yang, Narayan Prasad, and Xiaodong Wang.
\newblock A message-passing approach to distributed resource allocation in
  uplink dft-spread-ofdma systems.
\newblock {\em IEEE transactions on Communications}, 59(4):1099--1113, 2011.

\bibitem{bellalta2016ax}
Boris Bellalta.
\newblock {IEEE} 802.11 ax: High-efficiency {WLAN}s.
\newblock {\em IEEE Wireless Communications}, 23(1):38--46, 2016.

\bibitem{fwilhelmi2017code}
Francesc Wilhelmi.
\newblock Collaborative spatial reuse in wireless networks via selfih
  mulit-armed bandits.
\newblock
  \url{https://github.com/fwilhelmi/collaborative_sr_in_wns_via_selfish_mabs}.
  Commit: b9957e2, 2017.

\end{thebibliography}
	
\end{document}